\documentclass[prb,preprint,amsmath,amssymb,floatfix]{revtex4}

\usepackage{times}
\usepackage{amsmath}
\usepackage{graphicx}
\usepackage{dcolumn}
\usepackage{bm}
\usepackage{psfrag}
\usepackage{slashbox}

\begin{document}

\preprint{APS/123-QED}

\title{Transport properties of highly asymmetric hard sphere mixtures}

\author{Marcus~N.~Bannerman}
\author{Leo~Lue}
\email{leo.lue@manchester.ac.uk}
\affiliation{
School of Chemical Engineering and Analytical Science \\
The University of Manchester \\
PO Box 88 \\
Sackville Street \\
Manchester \\
M60 1QD \\
United Kingdom
}

\date{\today}

\begin{abstract}
  The static and dynamic properties of binary mixtures of hard spheres
  with a diameter ratio of $\sigma_B/\sigma_A=0.1$ and a mass ratio of
  $m_B/m_A=0.001$ are investigated using event driven molecular
  dynamics.  The contact value of the pair correlation functions are
  found to compare favourably with recently proposed theoretical
  expressions.  The transport coefficients of the mixture, determined
  from simulation, are compared to the predictions of revised Enskog
  theory, using both a third-order Sonine expansion and direct
  simulation Monte Carlo.  Overall, Enskog theory provides a fairly
  good description of the simulation data, with the exception of
  systems at the smallest mole fraction of larger spheres
  ($x_A=0.01$) examined.
  A ``fines effect'' was observed at higher packing fractions, where
  adding smaller spheres to a system of large spheres decreases the
  viscosity of the mixture; this effect is not captured by Enskog
  theory.
\end{abstract}

\pacs{Valid PACS appear here}


\maketitle

\section{Introduction}

Excluded volume interactions between molecules play a major role in
determining the structure and properties of most fluids and colloidal
systems.  The hard sphere model, which captures the essence of these
interactions, has played a central role in our understanding of the
properties of fluids, serving as a starting point of perturbation
theories for the description of real fluids
\cite{HANSEN_MCDONALD_1986}.  
%
Recently, there has been interest in binary hard sphere mixtures,
where the diameters of the two components are very different.  These
systems serve as models for nanoparticle suspensions and
colloid-polymer mixtures.
In these systems, an entropically driven depletion force
\cite{Oosawa_Asakura_1954,Vrij_1976} drives the larger particles to
cluster.
%
%
While there have been many studies on the structural (e.g., radial
distribution function) and thermodynamic properties (e.g., equation of
state) of these mixtures
\cite{JACKSON_ETAL_1987,Yau_etal_1996,Yau_etal_1997,HENDERSON_ETAL_1996_2,LUE_WOODCOCK_1999,Chan_Henderson_2000,LUE_WOODCOCK_2002,Henderson_etal_2005,Santos_2005,Vrabecz_Toacuteth_2006,Alawneh_Henderson_2008,Alawneh_Henderson_2008err},
there have been relatively few studies on their dynamical properties.
%

%


Much of the previous simulation work for the dynamical properties of
binary mixtures has focused on tracer particle studies
\cite{HERMAN_ALDER_1972,LEVITT_DAVIS_1974,ALDER_ALLEY_DYMOND_1974},
the velocity auto correlation functions, or the self-diffusion
coefficients
\cite{RUDYAK_KHARLAMOV_BELKIN_2000,RUDYAK_KHARLAMOV_BELKIN_2001}, as
these are relatively computationally inexpensive to determine.
These studies have revealed that the dynamics of the larger particles
deviates significantly from both the theoretical predictions of
Brownian particles and of Enskog theory.
Lue and Woodcock \cite{LUE_WOODCOCK_1999,LUE_WOODCOCK_2002} examined
the self-diffusion coefficients of size asymmetric binary mixtures of
hard spheres.  They found a ``fines effect'' at high densities, where
the addition of smaller spheres enhances the mobility of the larger
spheres.

Significantly less data are available for other dynamical properties.
Easteal and Woolf \cite{EASTEAL_WOOLF_1990} have investigated the
tracer diffusion coefficient for binary hard sphere mixtures. They
observe an inverse isotopic mass effect, where heavier tracer
particles diffuse faster beyond a certain solvent density than lighter
tracer particles.
Due to the computational cost of simulating highly size asymmetric
systems, past studies have focused on small size disparity and/or moderate
mole fractions of colloidal particles.

Erpenbeck \cite{ERPENBECK_1989,ERPENBECK_1992,ERPENBECK_1993} provided
the first complete transport study, comparing predictions from Enskog
theory and molecular dynamics results for binary hard sphere mixtures
approximating a Helium-Xenon gas mixture.  The mutual diffusion,
thermal diffusion, thermal conductivity and shear viscosity are given
over a range of state points.  Enskog theory was found to provide a
fairly good description of the transport properties for the conditions
studied.
Yeganegi and Zolfaghari \cite{Yeganegi_Zolfaghari_2006} have
investigated the thermal diffusion coefficient of binary hard spheres
(for moderate size ratios) using non-equilibrium molecular dynamics.
They observe a minimum in the thermal diffusion with density and good
agreement with Enskog theory.
%
Recently, Bastea \cite{Bastea_2007} has investigated the viscosity and
thermal conductivity of highly asymmetric ``soft-sphere'' mixtures at
very low volume fractions of the larger spheres.  Enskog theory was
only able to qualitatively describe the results in that study.

%

In the present work, we perform event driven molecular dynamics
simulations to study the static and transport properties of binary
hard sphere mixtures with a diameter ratio of 0.1 and a mass ratio of
0.001.  One of the motivations of this work is to further explore the
``fines effect'' revealed in these systems in a previous study by Lue
and Woodcock \cite{LUE_WOODCOCK_2002}.
Another aim of this work is to quantitatively test the predictive
ability of the revised Enskog theory \cite{DEHARO_ETAL_1983} for these
binary hard sphere systems over a broad range of conditions.
The remainder of this paper is organized as follows.  Details of the
hard sphere mixture model and the relation of the transport
coefficients to the microscopic dynamics of the system are discussed
in Section~\ref{sec:background}.  The details of the molecular
dynamics calculations and the direct simulation Monte Carlo solution
of the Enskog equation are provided in Section~\ref{sec:simulation}.
The simulation data for the static and the transport properties of the
binary hard sphere mixtures are presented in
Section~\ref{sec:results}, and the results are compared against the
predictions of the Enskog theory.
Finally, the main findings of this work are summarized in
Section~\ref{sec:conclusions}.

\section{Theoretical background
\label{sec:background}}

We consider systems consisting of additive hard spheres with differing
diameters and masses.  Spheres of type $a$ have a diameter $\sigma_a$
and a mass $m_a$.
The spheres are not permitted to overlap, and so the interaction
potential $u_{ab}$ between a sphere of type $a$ and a sphere of type
$b$ is given by
\begin{align}
u_{ab}\left(r\right) = \left\{
\begin{array}{ll}
  \infty & \mbox{if $r \le \sigma_{ab}$}\\        
  0 & \mbox{if $r > \sigma_{ab}$} \\
\end{array}
\right.
\end{align}
where $r$ is the distance between the centers of the two spheres, and
$\sigma_{ab}=\left(\sigma_a+\sigma_b\right)/2$.  
Due to the simple nature of this interaction potential, all properties
of hard sphere mixtures have a trivial dependence on the temperature.

One major advantage of the hard sphere model is the simplicity of its
dynamics.  The dynamics of hard sphere systems is driven by collisions
between spheres.  Between collisions, the spheres travel at constant
velocity.
The solution of the trajectory of the system then reduces to
determining the sequence of collisions between the spheres.  These
collisions alter the velocities of the spheres but conserve their
energy and momentum.
After a collision between a sphere $i$ of type $a$ and a sphere $j$ of type $b$,
the velocities of the spheres become ${\bf v}_i'$ and ${\bf v}_j'$
\begin{equation}
\label{eq:collision_rule}
\begin{split}
{\bf v}_i' &= {\bf v}_i 
- \frac{2\mu_{ab}}{m_a}
  \left({\bf v}_{ij}\cdot\hat{\bf r}_{ij}\right)\hat{\bf r}_{ij}
\\
{\bf v}_j' &= {\bf v}_j 
+ \frac{2\mu_{ab}}{m_b}
  \left({\bf v}_{ij}\cdot\hat{\bf r}_{ij}\right)\hat{\bf r}_{ij}
\end{split}
\end{equation}
where ${\bf v}_i$ and ${\bf v}_j$ are the velocities of the spheres
immediately before collision, $\hat{\bf r}_{ij}$ is a unit vector
pointing from the center of sphere $i$ to the center of sphere $j$,
${\bf v}_{ij}={\bf v}_i-{\bf v}_j$ is their relative velocity, and
$\mu_{ab}=m_am_b/(m_a+m_b)$ is the reduced mass.

\subsection{Static properties}

The pair correlation functions give an indication of the average local
environment of the particles in a system.  For hard sphere systems,
the values of the pair correlation functions at contact
$g_{ab}(\sigma_{ab}^+)$ play an important role.
In particular, they are directly related to the collision rates
between the spheres:
\begin{align}
g_{ab}\left(\sigma_{ab}^+\right)=\left(4 \pi \rho_b \sigma^2_{ab}
  t_{ab}\right)^{-1}(2 \pi \beta \mu_{ab})^{1/2}
\end{align}%
where $\rho_b$ is the number density of spheres of type $b$,
$\beta=(k_B T)^{-1}$, $k_B$ is the Boltzmann constant, $T$ is the
absolute temperature, and $t_{ab}$ is the mean time between which a
sphere of type $a$ undergoes collisions with a sphere of type $b$.
The quantity $t_{ab}$ can be calculated from the number of $a$-$b$
collisions $N_{ab}^{\rm (coll)}$ that occur in a simulation of
duration $t$
\begin{align}
t_{ab} = \frac{N_a t}{2N_{ab}^{\rm (coll)}}
\end{align}%
where $N_a$ is the number of spheres of type $a$ in the system.  
An advantage of molecular dynamics simulations over Monte Carlo
simulations is that the contact values of the pair correlation
functions can be directly calculated from the times $t_{ab}$ and does
not require the extrapolation of the pair correlation to contact.
%
%

The contact values of the pair correlation functions are also directly
related to the equation of state of the hard sphere system:
\begin{align}
\label{eq:virialEOS}
\frac{\beta p}{\rho} 
= 1 + \frac{2 \pi \rho}{3} \sum_{a,b} x_a x_b \sigma_{ab}^3 
  g_{ab}\left(\sigma_{ab}^+\right)
\end{align}%
where $p$ is the system pressure, $\rho$ is the total number
density of spheres, $x_a$ is the mole fraction of spheres of type $a$,
and the lowercase Latin indexes run over all species (i.e.\ $A$ and
$B$ for a binary mixture) present in the system.

Due to the fundamental importance of the contact values of the pair
correlation functions for hard sphere systems, there have been many
efforts to develop expressions to describe them
\cite{Yau_etal_1996,Yau_etal_1997,Chan_Henderson_2000,Matyushov_Ladanyi_1997}.
%
%
One of the most popular is the Boublik-Mansoori-Carnahan-Starling
(BMCSL) equation of state
\cite{Boublik_1970,MANSOORI_CARNAHAN_STARLING_1971}, which is an
interpolation between the virial and compressibility expressions of
the Percus-Yevick theory \cite{PERCUS_YEVICK_1958}.
%
%
This is given by
\begin{align} 
g_{ab}^{\rm BMCSL}\left(\sigma_{ab}^+\right)&= \frac{1}{1-\xi_3} 
+ \frac{3\xi_2}{2(1-\xi_3)^2}\frac{\sigma_a\sigma_b}{\sigma_{ab}}
+ \frac{\xi_2^2}{2(1-\xi_3)^3}\frac{\sigma_a^2\sigma_b^2}{\sigma_{ab}^2}
\label{eq:BMCSLgr}
\end{align}%
where $\xi_n$ is defined by
\begin{equation}
\xi_n = \frac{\pi \rho}{6} \sum_a x_a \sigma_a^n
\end{equation}%
Note that the solid fraction occupied by the spheres is given by
$\phi=\xi_3$.

The BMCSL equation yields predictions that are generally in good
agreement with simulation data for hard sphere mixtures over a broad
range of diameters and compositions \cite{JACKSON_ETAL_1987}.
However, for highly size asymmetric binary systems at small mole
fractions of the larger spheres (often referred to as the colloidal
limit), the BMCSL significantly underpredicts the contact value of the
pair correlation function between the larger spheres, as compared to
simulation results
\cite{JACKSON_ETAL_1987,LUE_WOODCOCK_1999,Roth_etal_2000}.

Recently, there have been several efforts to correct this.  Viduna and
Smith \cite{Viduna_Smith_2002,Viduna_Smith_2002_2} have suggested a
new expression, based on an empirical equation of state
\begin{align}
  g_{ab}^{\rm VS}\left(\sigma_{ab}^+\right)&= \frac{1}{1-\xi_3} +
  \frac{3-\xi_3+\xi_3^2/2}{2(1-\xi_3)^2}\xi_2
  \frac{\sigma_a\sigma_b}{\sigma_{ab}}
  +\frac{2-\xi_3-\xi_3^2/2}{6(1-\xi_3)^3}(2\xi_2^2+\xi_1\xi_3)
  \frac{\sigma_a^2\sigma_b^2}{\sigma_{ab}^2}
\end{align}%
This compact expression appears to compare well with simulation
results.  In the case of binary hard sphere mixtures, Henderson et
al.\ \cite{Henderson_etal_2005} have suggested further modifications
to the BMCSL and VS equations so that the contact value of the pair
correlation function between the larger spheres yield the correct
limiting behavior as the diameters of the larger spheres become
infinite \cite{Roth_etal_2000}.  Their expressions for the
pair correlation functions (which we denote as HC2) are given by
\begin{align}
  g_{BB}^{\rm HC2}\left(\sigma_{ab}^+\right) 
  &= g_{BB}^{\rm BMCSL}\left(\sigma_{BB}^+\right)
\text{\ \ or\ \ } g_{BB}^{\rm VS}\left(\sigma_{BB}^+\right)
\label{eq:HC2BB}
\\
  g_{AB}^{\rm HC2}\left(\sigma_{ab}^+\right) 
  &= g_{AB}^{\rm BMCSL}\left(\sigma_{AB}^+\right) 
  + \frac{\xi_2^2\sigma_{BB}^2}{\left(1-\xi_{3}\right)^3}
  \frac{1-R^2}{\left(1+R\right)^2}
  -\frac{\xi_2^3 \sigma_{BB}^3}{\left(1-\xi_3\right)^3}
  \frac{1-R^3}{\left(1+R\right)^3}
\label{eq:HC2AB}
\\
  g_{AA}^{\rm HC2}\left(\sigma_{ab}^+\right) 
  &= g_{AA}^{\rm VS}\left(\sigma_{BB}^+\right)
  +e^x-1-x-x^2/2
  \label{eq:HC2AA}
\end{align}%
where $A$ refers to the larger spheres, $B$ refers to the smaller
spheres, $R=\sigma_B/\sigma_A$ is the diameter ratio, and
$x=3\left(\xi_2\sigma_{AA}-\xi_3\right)/2$.  


\subsection{Calculation of transport coefficients
\label{sec:transport}}

In the continuum description of fluids \cite{GROOT_MAZUR_1984},
balance equations are typically used to relate the conserved
properties of the system (e.g., energy, momentum, and mass) to their
fluxes.  To close these equations, constitutive relations are
required.  These relations link the diffusive fluxes to gradients in
the thermodynamic properties of the system.  Transport coefficients
are defined through the assumption that the diffusive fluxes depend
linearly on the thermodynamic driving forces, which are gradients of
local thermodynamic properties of the system.
%

There are several possible choices \cite{GROOT_MAZUR_1984} for the
thermodynamic forces ${\bf X}$ and the diffusive fluxes ${\bf J}$.
For NVE molecular dynamics simulations, the most convenient
\cite{ERPENBECK_1989} choice is the ``mainstream'' (or ``unprimed''
\cite{GROOT_MAZUR_1984,ERPENBECK_1989}) definition of the fluxes.
These are defined as
\begin{subequations}\label{eq:TransportDefs}
\begin{align}
{\bf X}_a &= - T \bm{\nabla}\left(\frac{\mu_a}{T}\right) &
{\bf X}_\lambda &= - \frac{1}{T}\bm{\nabla}T 
\\
{\bf J}_a &= L_{a\lambda}{\bf X}_\lambda  + \sum_b L_{ab}{\bf X}_b &
{\bf J}_\lambda &= L_{\lambda\lambda} {\bf X}_\lambda
  + \sum_a L_{\lambda a} {\bf X}_a 
\end{align}%
\end{subequations}%
where $\mu_a$ is the chemical potential, and ${\bf J}_a$ is the
diffusive flux of species $a$, ${\bf J}_\lambda$ is the energy flux,
$L_{\lambda\lambda}$ is the thermal conductivity, $L_{ab}$ is the
mutual diffusion coefficient, and $L_{a\lambda}$ is the thermal
diffusivity.
The transport coefficients are defined through
Eqs.~(\ref{eq:TransportDefs}).


The relationship between stress tensor $\bm{\tau}$ and the strain rate
in the fluid is defined in the standard manner:
\begin{align}
\bm{\tau} &= p{\bf 1}
+ \left(\frac{2}{3}\eta-\kappa\right)
  \left(\bm{\nabla}\cdot{\bf u}\right) {\bf 1}
- \eta \left[\bm{\nabla}{\bf u}+\left(\bm{\nabla}{\bf u}\right)^T\right]
\end{align}%
where $\eta$ is the shear viscosity, $\kappa$ is the bulk viscosity,
and ${\bf u}$ is the streamline velocity of the fluid.  The quantity
${\bf{}1}$ represents the unit matrix, and the superscript ${}^T$
indicates the transpose of a matrix.

The Onsager reciprocity relations ($L_{ab}=L_{ba}$ and
$L_{a\lambda}=L_{\lambda a}$), combined with the requirement that
$\sum_a {\bf J}_a=0$ (due to the definition of the diffusive flux)
which implies $L_{aa}=-\sum_{b\neq{}a}L_{ab}$, reduce the number of
independent transport coefficients to $L_{\lambda\lambda}$,
$L_{A\lambda}$, $L_{AA}$, $\eta$, and $\kappa$.
In the following section, we discuss how these transport coefficients
can be determined from equilibrium molecular dynamics simulations.

\subsection{Einstein forms of the Green-Kubo relations}


The Green-Kubo formulas relate the time correlation functions of the
microscopic fluxes directly to the transport coefficients
\cite{HANSEN_MCDONALD_1986}. However, the Green-Kubo relations are an
unpopular method for obtaining the transport coefficients from
molecular dynamics simulations, as they require long simulation times
to obtain good statistics.  This is not a significant issue in hard
sphere systems, as long simulation times are more easily
accessible. For systems with particles interacting with discontinuous
potentials, the Einstein form of the Green-Kubo relations must be
used, due to the impulsive nature of the interaction potential. The
full derivation of the these formulas are already available
\cite{ERPENBECK_1989,HANSEN_MCDONALD_1986}, and, therefore, only the
final expressions are presented here for completeness.

\begin{figure}[ht]
\begin{center}
\includegraphics[clip,width=\columnwidth]{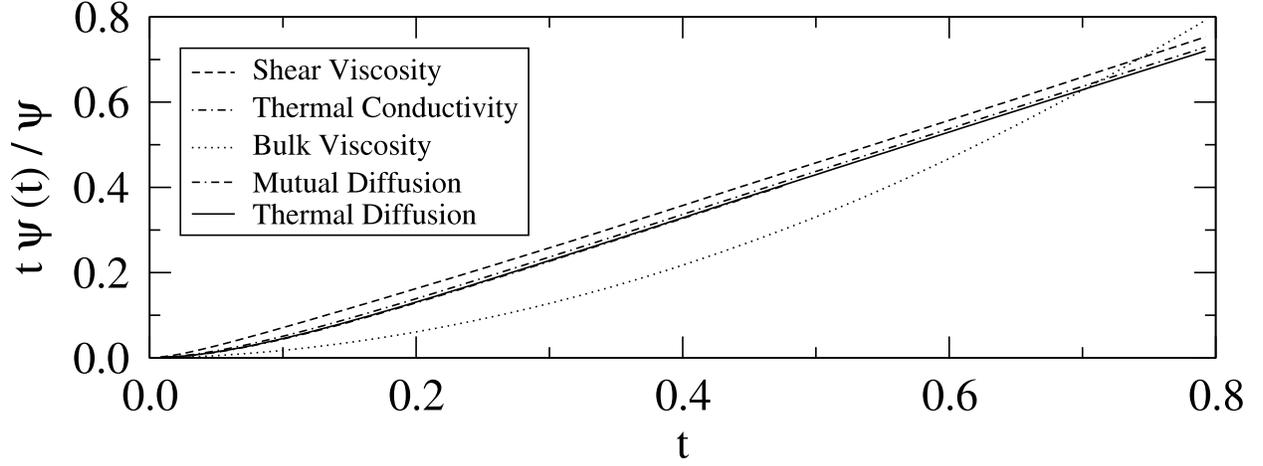}
\caption{\label{fig:CorrSample} 
  Time dependent transport coefficients (see Eq.~(\ref{eq:Einstein})),
  reduced by their infinite time result, from a single simulation run
  for a binary hard sphere system with $x_A=0.01$ and solid fraction
  $\phi=0.1$.  The time is presented in units of $(\beta
  m_A\sigma_A^2)^{1/2}$; the mean free time is roughly $0.015(\beta
  m_A\sigma_A^2)^{1/2}$.}
\end{center} 
\end{figure}

The Einstein relations have the general form
\begin{align}
\label{eq:Einstein}
\psi(t) =  \frac{\beta}{2 V t}
\left\langle W_{\psi_1}(t) W_{\psi_2}(t)\right\rangle
\end{align}%
where $\psi(t)$ is a time dependent transport coefficient, $V$ is the
volume of the system, and $W_{\psi 1}$ and $W_{\psi 2}$ are
displacement functions corresponding to time integrals of the
microscopic fluxes.  The displacement functions for a system with zero
total momentum in the microcanonical ensemble are given in
Table~\ref{tab:displacementfunc}.
The pair of displacement functions that correspond to each of the
transport coefficients are summarized in Table~\ref{tab:coeffs}.
In hydrodynamic regime, the transport coefficients are given by the
infinite time limit of Eq.~(\ref{eq:Einstein})
\begin{align}
\psi &= \lim_{t\to\infty} \psi(t)
\end{align}%
A sample of reduced correlators for a single molecular dynamics
simulation run is plotted in Fig.~\ref{fig:CorrSample}.  The function
$t\psi(t)$ typically displays transient behavior for short times
before changing to the linear, long-time regime.  All the transport
properties, with the exception of the bulk viscosity, rapidly
transition to the linear regime within a few mean free times.  The
bulk viscosity, however, only slowly approaches the linear regime,
and, consequently, the limiting values are difficult to extract.  As a
result, we do not present data for the bulk viscosity.

A time correlation function of a finite sized simulation is only
representative of a bulk system for a limited duration. Beyond the
time a sound wave takes to traverse the simulation box, the system
size begins to affect the correlation function. The sound wave
traversal time is determined directly from the speed of sound,
$c$. For a hard sphere system the speed of sound is given by
\begin{align}
  c^2 &= m^{-1} k_BT \left[\frac{2 Z^2}{3}
    +\frac{\partial \rho Z}{\partial \rho}\right]
  \label{eq:speedofsound}
\end{align}%
where $Z=\beta p / \rho$ is the compressibility factor, and $m=\sum_a
x_a m_a$ is the mean particle mass.  The HC2 equation of state (see
Eqs.~(\ref{eq:virialEOS}), (\ref{eq:HC2BB}), (\ref{eq:HC2AB}), and
(\ref{eq:HC2AA})) is used to estimate the speed of sound, via
Eq.~\eqref{eq:speedofsound}.  Data for the time correlation functions
are only collected for a duration of time shorter than the sound wave
traversal time.

\subsection{Enskog theory predictions for the transport coefficients}

Revised Enskog theory (RET)
\cite{DEHARO_ETAL_1983,KINCAID_ETAL_1983,DEHARO_COHEN_1984,KINCAID_ETAL_1987}
is an extension of the highly successful Enskog theory to mixtures.
This is the most widely applied kinetic theory of moderately dense
fluids.  In the Enskog approximation, all pre-collision correlations
between particles are ignored, save for a single static structural
correlation function.  In a homogeneous system, this reduces to the
values of the various pair correlation functions at contact, which
govern the collision rates.  Given these as input, Enskog theory
yields predictions for the transport properties through the
Chapman-Enskog expansion \cite{CHAPMAN_COWLING_1970}.

The standard method to solve to the Enskog equation is to expand the
one-particle distribution function in a series of Sonine polynomials.
Erpenbeck \cite{ERPENBECK_1989} has compiled the (corrected) Enskog
expressions for all transport properties, excluding the bulk
viscosity, of hard sphere mixtures.  These expressions have been
combined with the table of integrals given by Ferziger and
Kaper~\cite{FERZIGER_KAPER_1972} and a linear equation solver to
evaluate Enskog theory to the third order in the Sonine expansion.  We
present results calculated from the BMCSL and HC2 equations to
determine the effect of improved values for $g_{ab}(\sigma_{ab}^+)$ on
the predictions of the transport properties.

\subsection{DSMC solution of the Enskog equation}


Another method for obtaining solutions to the Enskog equations is
through the use of the direct simulation Monte Carlo (DSMC) method.
This technique was originally developed for the Boltzmann equation but
has recently been extended to the Enskog equation
\cite{Montanero1996a,Montanero1997,Frezzotti_1997}.  In this work,
DSMC of the Enskog equation, in the style of Bird's NTC method
\cite{bird_1994}, is used to provide results. In this approach, the
velocity distribution of each species is approximated using a set of
samples
\begin{align}
  f_a({\bf v}, t) = {\mathcal N}_{a}^{-1} 
\sum_{i=1}^{{\mathcal N}_{a}} 
\delta\left({\bf v} - {\bf v}_i\left(t\right)\right)
\end{align}
where ${\mathcal N}_{a}$ is the number of samples of the velocity
distribution of species $a$.  For simplicity, in the following
expressions we assume each sample represents a single sphere.  Other
choices are possible; however, the difference merely affects the
relative sample collision testing rates and time scale of the simulation.

The probability that a sample $i$ of species $a$ undergoes a collision
event with species $b$ after a time step $\Delta t_{ab}$ is
\cite{Hopkins_Shen_1992}
\begin{align}
  \label{eq:collprob}
  \omega_{ib}=  4 g_{ab}\left(\sigma_{ab}^+\right) \pi \rho_{b} \sigma_{ab}^2
  \left({\bf v}_{ij}\cdot \hat{\bf k}\right) 
    \Theta\left({\bf v}_{ij}\cdot\hat{\bf k}\right) \Delta t_{ab}
\end{align}
where $j$ is a randomly chosen sample from species $b$, $\hat{\bf k}$
is a randomly chosen relative orientation between the samples on
collision, ${\bf{}v}_{ij}={\bf{}v}_i-{\bf{}v}_j$ is the relative
velocity, and $\Theta$ is the Heaviside step function.
The time step $\Delta{}t_{ab}$ describes the rate at which samples in
species $a$ are tested for collisions with a sample of species $b$.
For a DSMC calculation of a binary mixture, there are four rates, one for
each pairing of the species ($AA$, $AB$, $BA$, and $BB$).

The simplest DSMC algorithm proceeds by incrementing time to the next
test for collisions between species $a$ and $b$.  Each sample $i$ of
species $a$ is tested for an event with another sample $j$, which is
randomly selected.  A collision is executed with a probability given
by Eq.~\eqref{eq:collprob}.  This collision only affects sample $i$
and not the collision partner $j$.  This method is simple but
inefficient because properties that are conserved on collision (e.g.,
momentum and energy) are only conserved on average. In addition, all
samples in species $a$ are tested at each time step, which is
computationally expensive, even though $\Delta{}t_{ab}$ is selected to
yield only a few events per time step.

An improved algorithm, based on Bird's NTC method, executes symmetric
species-species collision events simultaneously, and, therefore, there
are three independent test rates for the binary system ($\Delta
t_{AA}$, $\Delta t_{AB}=\Delta t_{BA}$, and $\Delta t_{BB}$).  
For a given time step, we assume there are a maximum of
$N^{pairs}_{ab}={\mathcal N}_{a}\omega_{ab}^{(max)}={\mathcal
  N}_{b}\omega_{ba}^{(max)}$ events that may occur for each species;
the quantity $\omega_{ba}^{(max)}$ is the maximum observed value of
$\omega_{ba}$, which is updated, if required, during the course of a
simulation. $N^{pairs}_{ab}$ pairs of $a$ and $b$ samples are randomly
selected at each time step.  The probability of collision is then
scaled to
\begin{align}
\frac{1}{2} (2-\delta_{ab}) \omega_{ab}
\frac{{\mathcal N}_{a}}{N_{pairs}}
\end{align}
where $\delta_{ab}$ is the Kronecker delta.  If the collision is
accepted, then the velocities of both samples are updated according to
the collision rule (see Eq.~(\ref{eq:collision_rule})).  This
conserves energy and momentum at all times and greatly improves the
statistics of the simulation.  Like Enskog theory, the DSMC
calculations require $g_{ab}\left(\sigma_{ab}^+\right)$ as input,
however, DSMC requires no polynomial expansion to make the problem
tractable. 

The transport coefficients are obtained through the use of the
appropriate time correlation functions, as in the full molecular
dynamics simulations (see Section~\ref{sec:transport}).  DSMC provides
an attractive method of numerically solving a kinetic equation,
especially as computing power increases.  Its results are still,
however, limited by the approximations of the underlying kinetic
equation.

\section{Simulation details
\label{sec:simulation}}

In this work, we examine the static and transport properties of highly
asymmetric binary hard sphere mixtures.  The larger $A$ spheres have a
diameter $\sigma_A$ and mass $m_A$, and the smaller $B$ spheres have a
diameter $\sigma_B$ and mass $m_B$.  We consider systems with
$\sigma_B/\sigma_A=0.1$ and $m_B/m_A=0.001$, consistent with particles
of the same density.

Discrete potentials, such as the hard sphere model, have an important
advantage over more complex ``soft'' potentials.  Between collisions
the spheres or molecules experience no forces and travel on ballistic
trajectories.  The dynamics can be solved analytically, and the
integration of the equations of motion is processed as a sequence of
events.  Current event driven molecular dynamics algorithms are now
quite advanced and allow the simulation of large systems for the long
times required to extract accurate transport properties.

\subsection{MD Simulations}

The basic event driven algorithm used in this work to perform the
molecular dynamics simulations is fundamentally the same as the one
originally described by Alder and Wainwright
\cite{ALDER_WAINWRIGHT_1959}.  Neighbor lists and the delayed states
algorithm \cite{Marin_etal_1993} are included to optimize the
calculations.  These methods are combined with a new bounded priority
queue, suggested by Paul \cite{Paul_2007}, to remove the system size
dependence of sorting the event queue.  Finally, the interactions
between the largest spheres are removed from the neighbor list and
processed separately \cite{Vrabecz_Toacuteth_2006} to allow the use of
a smaller cell size and reduced number of collision tests.  This
removal is restricted to low mole fractions of the larger spheres as
the overhead of these removed interactions is of order $O(N^2)$ in the
number of large spheres.




A total of $N=13500$ spheres in a cubic box of volume $V$ with
standard periodic boundary conditions were used in all the
simulations.  The volume of the system and the relative number of
large and small spheres (i.e., $N_A$ and $N_B$) were adjusted to
obtain the required packing fraction and composition, respectively.
For each of the systems examined, the initial configurations were
equilibrated over a period of $10^7$ collisions and then run for $20$
trajectories of $10^8$ collisions to collect the collision statistics
and time correlation functions.

The time correlation functions for the various transport properties
were collected over approximately $100$ intervals of a mean free time,
using the start time averaging method \cite{Haile_1997}.  The last
$50$ values of the correlator were fitted to a line to extract the
long time limit of the transport coefficient.

\subsection{DSMC simulations}

DSMC simulations were performed using a total of ${\mathcal
  N}_{A}+{\mathcal N}_{B}=13500$ samples of the velocity distribution.
Each of the simulations was initially equilibrated for $10^7$
collisions.  The time correlation functions were then collected over
$8$ separate trajectories, each consisting of $10^8$ collisions, using
$100$ intervals of a mean free time.  The statistical uncertainty of
the shorter DSMC calculations are smaller than the uncertainties of
the MD simulations because Enskog theory neglects dynamical
correlations.

\section{Results and discussion
\label{sec:results}}

In this section, we present the results of the molecular dynamics
simulations for the contact value of the pair correlation functions
and the transport coefficients of binary hard sphere mixtures.  A
comparison of the predictions of the revised Enskog theory is also
provided.
All quantities are reported in reduced units, where the unit of mass
is $m_A$, the unit of length is $\sigma_A$, and the unit of energy is
$k_BT$.

\subsection{Static properties}

\begin{figure}[ht]
\begin{center}
\includegraphics[clip,width=\columnwidth]{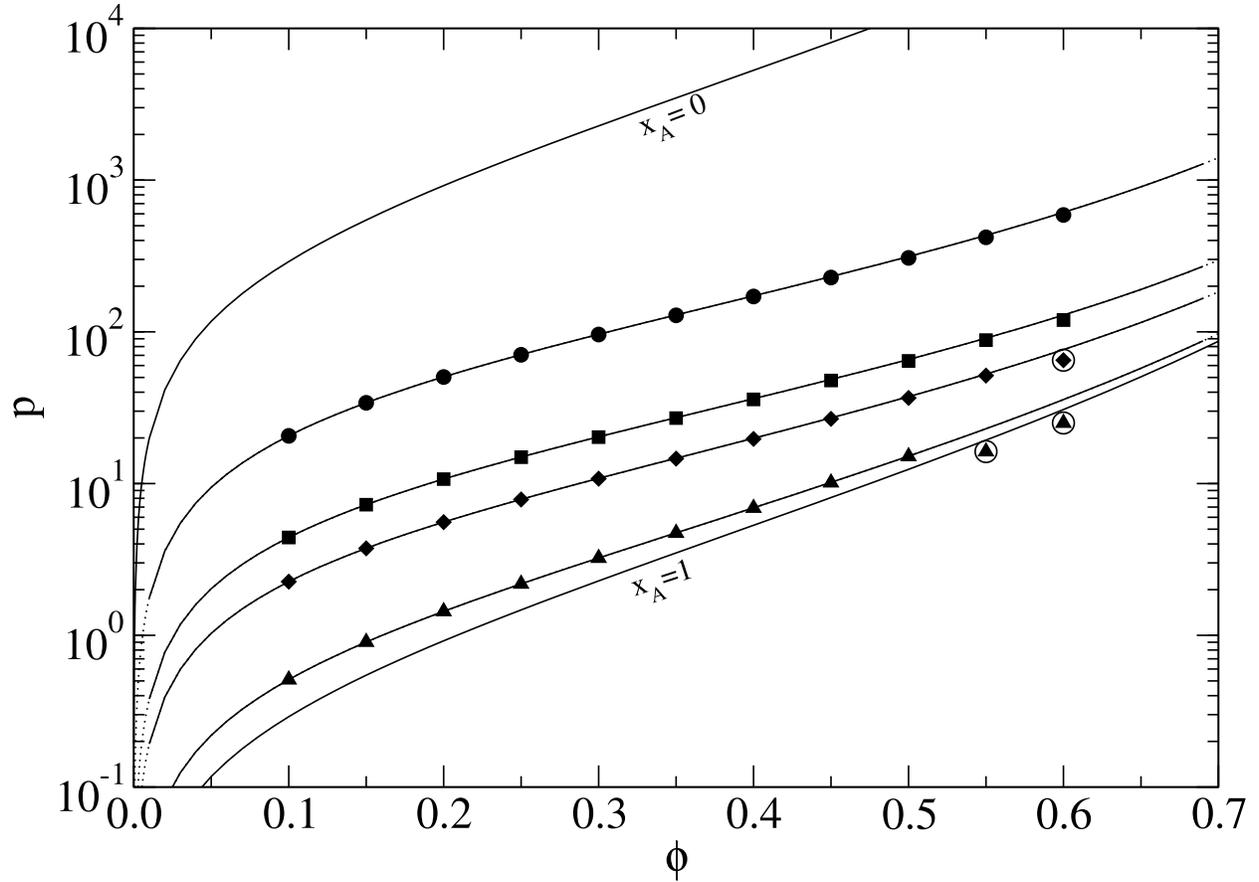}
\caption{\label{fig:Pressure} 
Pressure $p$ as a function of solid
fraction $\phi$ for binary hard sphere mixtures with
$\sigma_B/\sigma_A=0.1$, $m_B/m_A=0.001$, and
(i) $x_A=0.01$ (circles),
(ii) $x_A=0.05$ (squares), 
(iii) $x_A=0.1$ (diamonds), and
(iv) $x_A=0.5$ (triangles).
The filled symbols are from molecular dynamics simulations, the lines
are the predictions of the BMCSL (solid) and HC2 (dotted) equations of
state. Data points are circled where the system shows signs of freezing.}
\end{center} 
\end{figure}

The variation of the pressure of the binary hard sphere mixtures with
packing fraction and composition is shown in Fig.~\ref{fig:Pressure}.
The symbols are the data from the molecular dynamics simulations, and
the lines are the predictions of the BMCSL (solid) and HC2 (dotted)
equations of state.  These equations of state provide an excellent
description of the simulation data, with the exception of the very
highest packing fractions where they overpredict the pressure.  These
deviations, however, are due to the onset of freezing of the larger
spheres; the single component hard sphere fluid begins to freeze at a
packing fraction of $0.494$ \cite{Hoover_Ree_1968}.


The contact values of the $AA$, $AB$, and $BB$ pair correlation
functions are plotted in Fig.~\ref{fig:chi} as a function of the total
volume fraction of spheres for different mole fractions of the larger
$A$ spheres $x_A$.
The simulation results for $g_{BB}$ are well described by the BMCSL
theory.  This is in agreement with previous simulation studies of
binary hard spheres mixtures
\cite{LUE_WOODCOCK_1999,Alawneh_Henderson_2008}. The VS predictions
(not shown) provide equally accurate predictions for $g_{BB}$.

The BMCSL predictions for $g_{AB}$ lie above the simulation results at
high density for the lowest mole fraction studied. The HC2 predictions
are higher still, however, the error is within a few percent. The
corrections of Henderson et al.\ \cite{Henderson_etal_2005} to
$g_{AB}$ are small for the systems studied.  The VS predictions (not
shown) lie between the HC2 and the BMCSL results

For the contact value of pair correlation function between the larger
spheres, the BMCSL predictions fall significantly below the simulation
results at high density for the lowest mole fraction studied.  The HC2
predictions are exceptionally accurate, even for the smallest mole
fractions of the larger spheres.  This is due to the success of the
underlying VS equation (not shown), which give results that are nearly
indistinguishable from the HC2 equation.
At $\phi\approx0.55$, $g_{AA}\left(\sigma^+_{AA}\right)$ for the
$x_A=0.5$ system decreases significantly.  This also occurs in the
$x_A=0.1$ system at a higher packing fraction of $\phi=0.6$.  It
appears that the larger component has frozen while the smaller spheres
remain fluid.  

Overall, the HC2 expression is accurate and provides excellent
estimates for the contact values of the pair correlation functions for
all the conditions studied in this work.


\subsection{Thermal conductivity}

\begin{figure}[ht]
\begin{center}
\includegraphics[clip,width=\columnwidth]{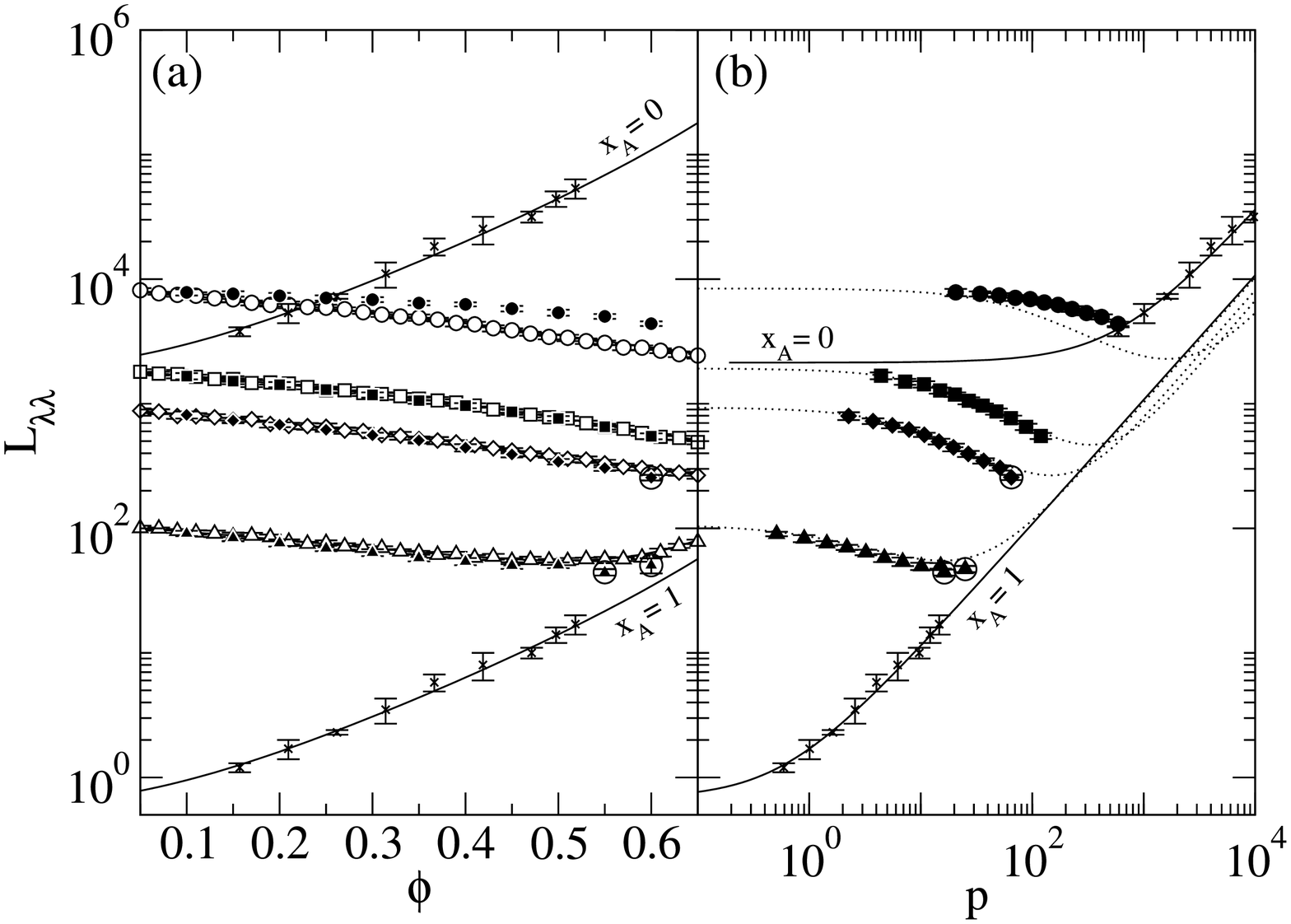}
\caption{\label{fig:thermcond} 
Thermal conductivity $L_{\lambda\lambda}$ as a function of solid
fraction $\phi$ (a) and pressure $p$ (b) for binary hard sphere mixtures with
$\sigma_B/\sigma_A=0.1$, $m_B/m_A=0.001$, and
(i) $x_A=0.01$ (circles),
(ii) $x_A=0.05$ (squares), 
(iii) $x_A=0.1$ (diamonds), and
(iv) $x_A=0.5$ (triangles).
The filled symbols are from molecular dynamics simulations, and the
open symbols are the DSMC results for the Enskog theory. The crosses
are molecular dynamics simulations for single component hard spheres,
taken from Ref.~\onlinecite{Lue_2005}. The lines are third order
Enskog theory predictions using BMCSL (solid) and HC2 (dotted) values
of $g_{ab}(\sigma_{ab}^+)$. Simulation data points are circled where systems show
signs of freezing.}
\end{center} 
\end{figure}

The thermal conductivity of the binary hard sphere mixtures is plotted
in Fig.~\ref{fig:thermcond}a with respect to the packing fraction and
in Fig.~\ref{fig:thermcond}b with respect to the pressure.  The
molecular dynamics simulation data are given by the filled symbols.
The crosses are molecular simulation data for single component hard
spheres, taken from Ref.~\onlinecite{Lue_2005}.
For single component hard sphere systems, the thermal conductivity
increases with increasing packing fraction and pressure.  The initial
addition of smaller spheres to a system of larger spheres (i.e.\
decreasing $x_A$) significantly increases the thermal conductivity of
the mixture.
At the same packing fraction, a system with a lower mole fraction of
larger spheres will have many more particles than a system with a
higher mole fraction of larger spheres.  These additional particles
enhance the ability of system to transport energy.
With the addition of smaller spheres to the large sphere system, we
observe that the thermal conductivity no longer increases
monotonically with the packing fraction (or the pressure).  Rather,
the thermal conductivity initially decreases with increasing packing
fraction down to a minimum value, and then it increases.  The packing
fraction at the minimum increases as the fraction of smaller spheres
increases.

Interestingly, at packing fractions below $\phi\approx0.25$, the
thermal conductivity of pure $B$ spheres (i.e., $x_A=0$) is lower than
the thermal conductivity for the $x_A=0.01$ system, while for
$\phi>0.25$ it is higher.  This implies that at sufficiently low
packing fraction (or pressure) the thermal conductivity of the system
must have a maximum with respect to $x_A$.  Physically, this would
correspond to a situation where the addition of larger spheres to a
fluid of smaller hard spheres would enhance its thermal conductivity.



\begin{figure}[ht]
\begin{center}
\includegraphics[clip,width=\columnwidth]{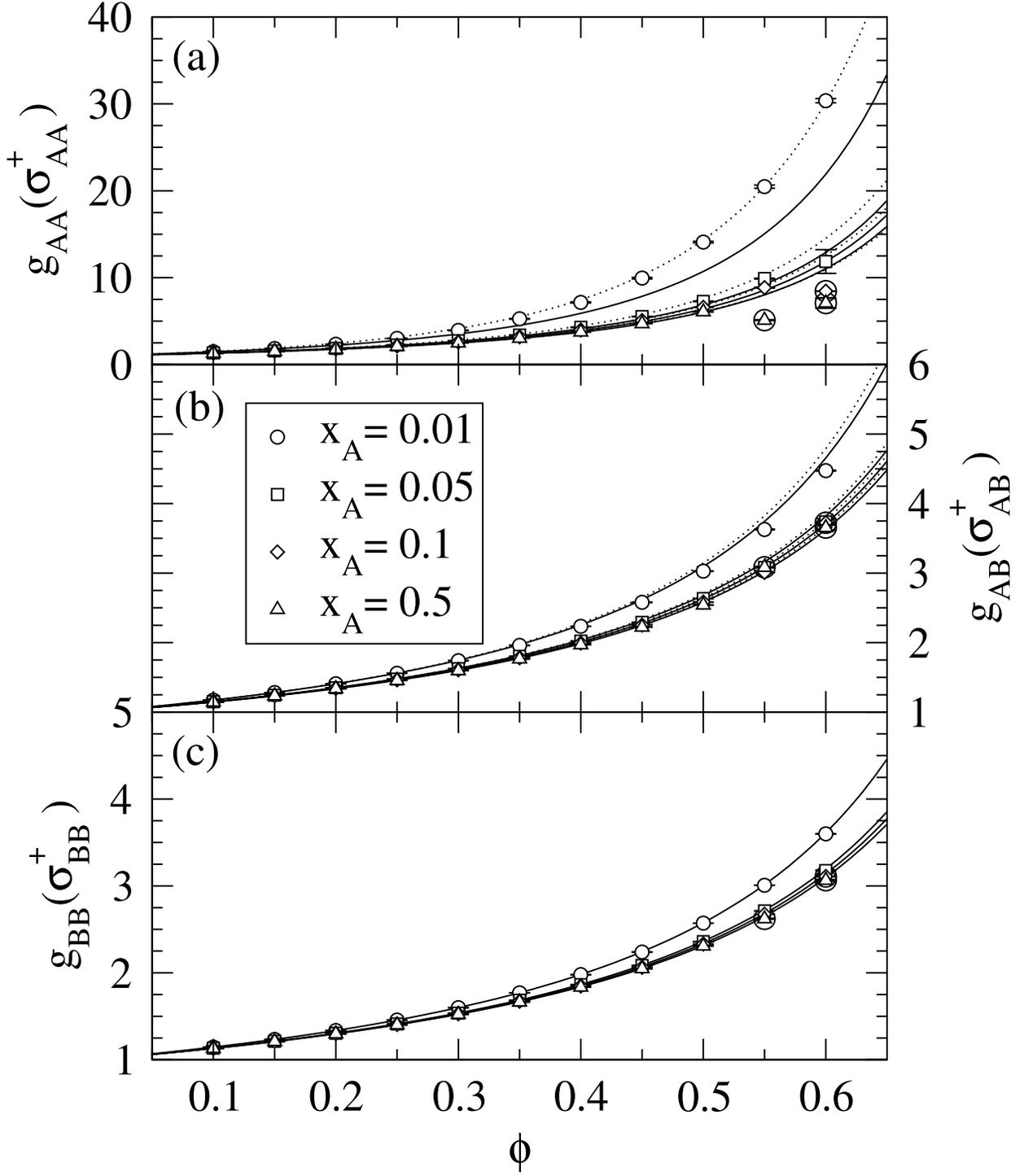}
\caption{\label{fig:chi} 
Contact value of the pair correlation function
$g_{ab}\left(\sigma_{ab}^+\right)$ between the large-large (a),
large-small (b), and small-small (c) sphere species as a function of
solid fraction $\phi$ for binary hard sphere mixtures with
$\sigma_B/\sigma_A=0.1$, $m_B/m_A=0.001$, and
  (i) $x_A=0.01$ (circles),
  (ii) $x_A=0.05$ (squares), 
  (iii) $x_A=0.1$ (diamonds), and
  (iv) $x_A=0.5$ (triangles).
  The solid lines are the predictions of the BMCSL equation (see
  Eq.~\eqref{eq:BMCSLgr}), and the dotted lines are the predictions of
  the HC2 equation (see Eq.~\eqref{eq:HC2AA}). Simulation data points
  are circled where the system shows signs of freezing.}
\end{center} 
\end{figure}

The solid lines in Fig.~\ref{fig:thermcond} are the predictions of
Enskog theory within the third order Sonine approximation with the
BMCSL expressions for the collision rates, while the dotted lines are
the third order Enskog predictions with the HC2 expressions.  The
difference between using the BMCSL and HC2 expressions in Enskog
theory is negligible, as the collisional contribution to the thermal
conductivity is dominated by the $BB$ and $BA$ interactions (see
Figs.~\ref{fig:chi}b and c).
The open symbols in Fig.~\ref{fig:thermcond} are from DSMC
calculations using the HC2 expressions for the collision rates.  These
results are nearly identical to the third order Sonine approximation,
indicating the accuracy of the approximation and validating the DSMC code.

The simulation results are well described by Enskog theory for the
pure hard sphere systems (i.e.\ $x_A=0$ and $1$), as well as for
mixtures with relatively high mole fractions of the larger spheres
($x_A\ge0.05$).  At high packing fractions, the Enskog predictions
deviate slightly for the case $x_A=0.5$; however, this occurs at the
conditions where component $A$ appears to freeze (see
Fig.~\ref{fig:chi}a), and the BMCSL and HC2 expressions for
$g_{ab}\left(\sigma_{ab}^+\right)$ are not applicable for solid
phases.

For $x_A=0.01$, Enskog theory significantly underpredicts the thermal
conductivity of the system.  This deviation may be related to the
enhanced mobility of the system due to the fines effect
\cite{LUE_WOODCOCK_2002} and is a result of a dynamic process not
captured by Enskog theory.
Note, however, that Enskog theory provides good predictions for the
thermal conductivity of one component hard sphere systems
\cite{Lue_2005}, so one expects that for vanishing amounts of the
larger spheres (i.e.\ the limit where $x_A\to0$), Enskog theory should
again provide a fairly good description of the simulation data.

\subsection{Shear viscosity}

\begin{figure}[ht]
\begin{center}
\includegraphics[clip,width=\columnwidth]{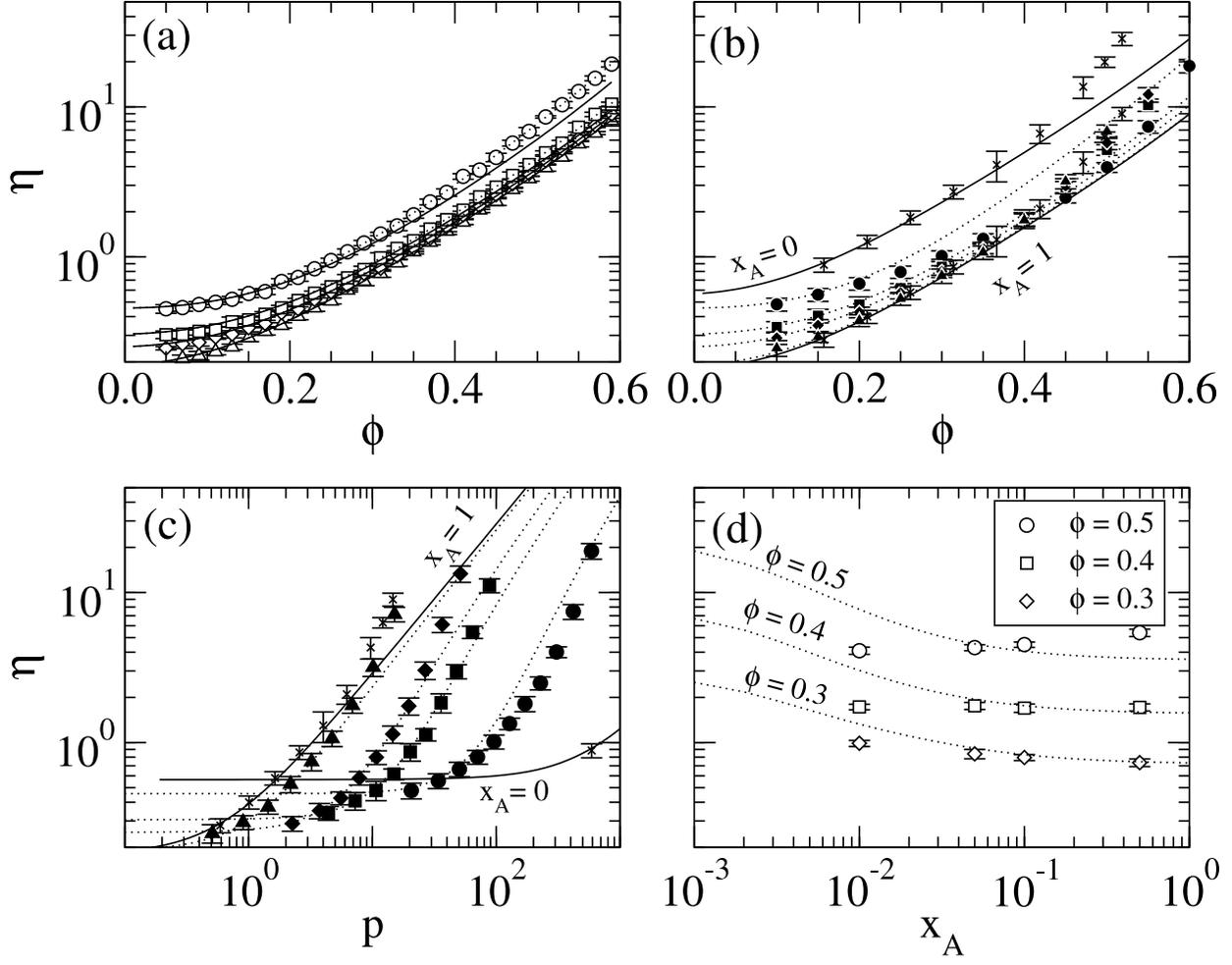}
\caption{\label{fig:shearvisc} 
  Shear viscosity $\eta$ as a function of solid fraction $\phi$ (a-b),
  pressure $p$ (c), and mole fraction $x_A$ (d) for binary hard sphere
  mixtures with \mbox{$\sigma_B/\sigma_A=0.1$} and
  \mbox{$m_B/m_A=0.001$}. With the exception of (d), the symbols
  indicate a mole fraction of
(i) $x_A=0.01$ (circles),
(ii) $x_A=0.05$ (squares), 
(iii) $x_A=0.1$ (diamonds), and
(iv) $x_A=0.5$ (triangles).
The filled symbols are from molecular dynamics simulations, and the
open symbols are the DSMC results for the Enskog theory. The crosses
are molecular dynamics simulations for single component hard spheres,
taken from Ref.~\onlinecite{Lue_2005}. The lines are third order
Enskog theory predictions using the BMCSL (solid) and HC2 (dotted)
predictions for $g_{ab}(\sigma_{ab}^+)$.}
\end{center} 
\end{figure}

The shear viscosity is plotted in Fig.~\ref{fig:shearvisc}.
The viscosity of all the mixtures increases monotonically with the
packing fraction of the spheres and the pressure of the system (see
Fig.~\ref{fig:shearvisc}a-c).
Unlike for the thermal conductivity, the Enskog theory predictions for
the shear viscosity using the HC2 expression for the collision rates
noticeably differ from the BMCSL results (see
Fig.~\ref{fig:shearvisc}a); however, this only occurs in regions where
Enskog theory poorly describes the simulation results (see
Fig.~\ref{fig:shearvisc}b and c).
%
Enskog theory captures the low density behavior of the viscosity quite
well.  For single component hard sphere systems, Enskog theory is
known to underpredict the viscosity at high densities
\cite{Lue_Bishop_2006}, due to its inability to account for correlated
collisions resulting from the caging of spheres at these conditions.
For the binary hard sphere mixtures that we study here, the Enskog
theory underpredicts the viscosity, in general.  However, the case
$x_A=0.01$ is an exception, where Enskog theory actually overpredicts
the viscosity at high packing fractions.

An interesting ``fines'' effect occurs in the viscosity of these
systems.  At low overall packing fractions (or pressures), the
addition of smaller spheres to a system of larger spheres (i.e.\
decreasing $x_A$) increases the viscosity of the system.  However,
above a packing fraction of about $\phi=0.4$, the curves for the
viscosity crossover, and the addition of smaller spheres to a system
of larger spheres {\em decreases} the viscosity of the system.  This
is highlighted in Fig.~\ref{fig:shearvisc}d where the viscosity is
almost independent of composition at a packing fraction of $\phi=0.4$.
The ``fines'' effect is not captured by Enskog theory, which indicates
its origin is in dynamical correlations between particles.  In these
systems, the presence of the smaller spheres leads to an attractive
depletion force \cite{Oosawa_Asakura_1954,Vrij_1976} between the
larger spheres, which is entropically driven.  This force may disrupt
the caging of larger spheres \cite{LUE_WOODCOCK_2002} by forcing them
into closer contact, thereby creating a more open network and
increasing the mobility of both species.


\subsection{Thermal diffusion coefficient}

\begin{figure}[ht]
\begin{center}
\includegraphics[clip,width=\columnwidth]{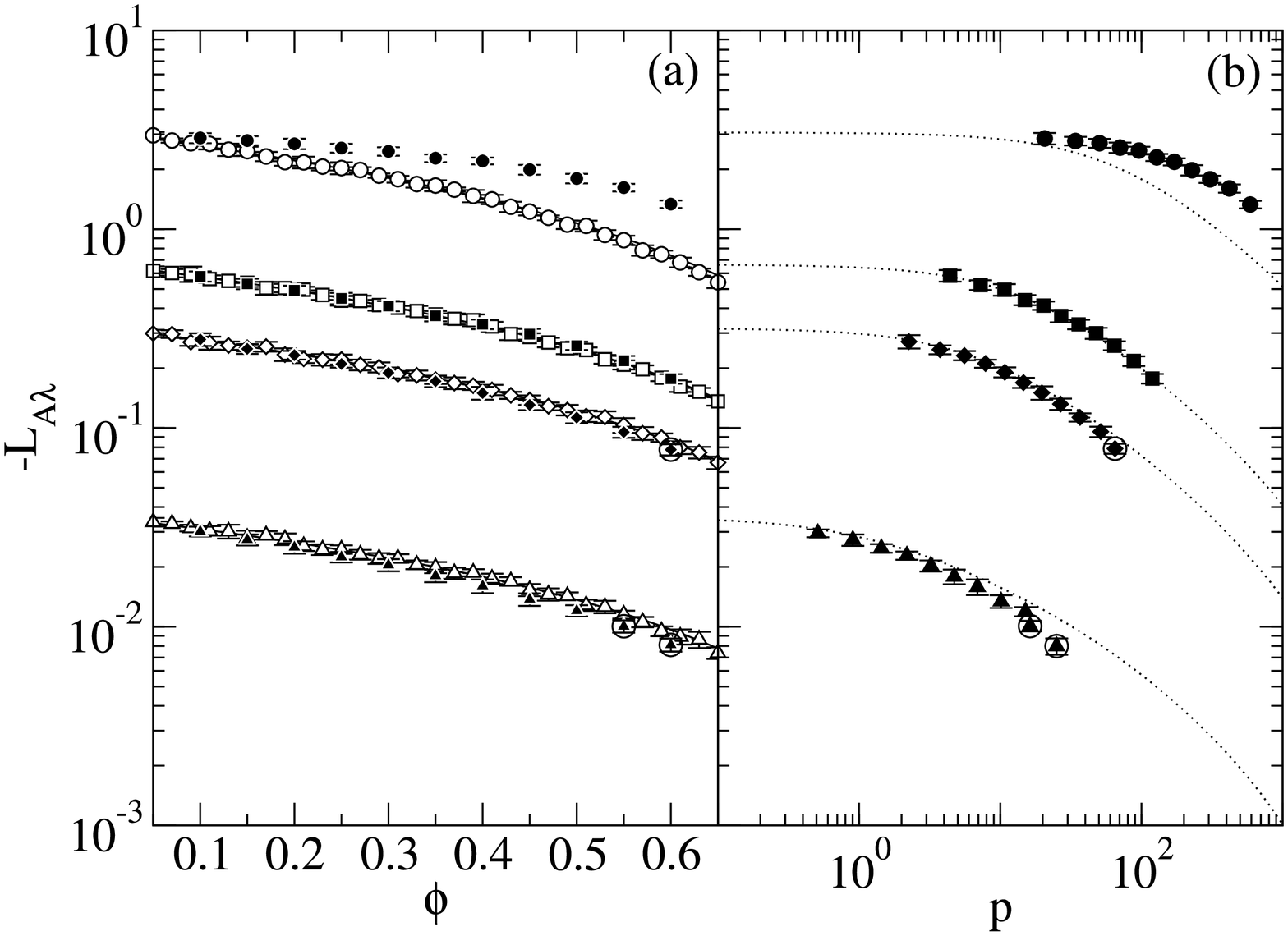}
\caption{\label{fig:thermaldiff} 
Thermal diffusivity $L_{A\lambda}$ as a function of solid fraction
$\phi$ (a) and pressure $p$ (b) for binary hard sphere mixtures with
\mbox{$\sigma_B/\sigma_A=0.1$}, \mbox{$m_B/m_A=0.001$}, and
  (i) $x_A=0.01$ (circles),
  (ii) $x_A=0.05$ (squares), 
  (iii) $x_A=0.1$ (diamonds), and
  (iv) $x_A=0.5$ (triangles).
  The filled symbols are from molecular dynamics simulations, and the
  open symbols are the DSMC results for the Enskog theory. The lines
  are third order Enskog theory predictions using the BMCSL (solid)
  and HC2 (dotted) predictions for $g_{ab}(\sigma_{ab}^+)$. Simulation
  data points are circled where the system shows signs of freezing.}
\end{center} 
\end{figure}

Figure~\ref{fig:thermaldiff} presents the thermal diffusivity of the
larger spheres over a range of packing fractions and pressures.
Because $L_{A\lambda}$ is negative, the larger species tends to move
towards regions of higher temperature.  Increasing the packing
fraction, the pressure, or the fraction of larger spheres in the
system decreases the magnitude of the thermal diffusivity.  This
general trend is in agreement with previous NEMD simulations
\cite{Yeganegi_Zolfaghari_2006}.

The use of the HC2 expressions with Enskog theory offers no
significant improvement on the BMCSL predictions, again due to the
dominance of the small spheres in the energy transport.  Enskog theory
is in quantitative agreement with the simulation data over a broad
range of conditions examined in this work.  However, the main
exception is for the composition $x_A=0.01$, where it substantially
underpredicts the $L_{A\lambda}$ at the higher packing fractions.

\subsection{Mutual diffusion coefficient}

\begin{figure}[ht]
\begin{center}
\includegraphics[clip,width=\columnwidth]{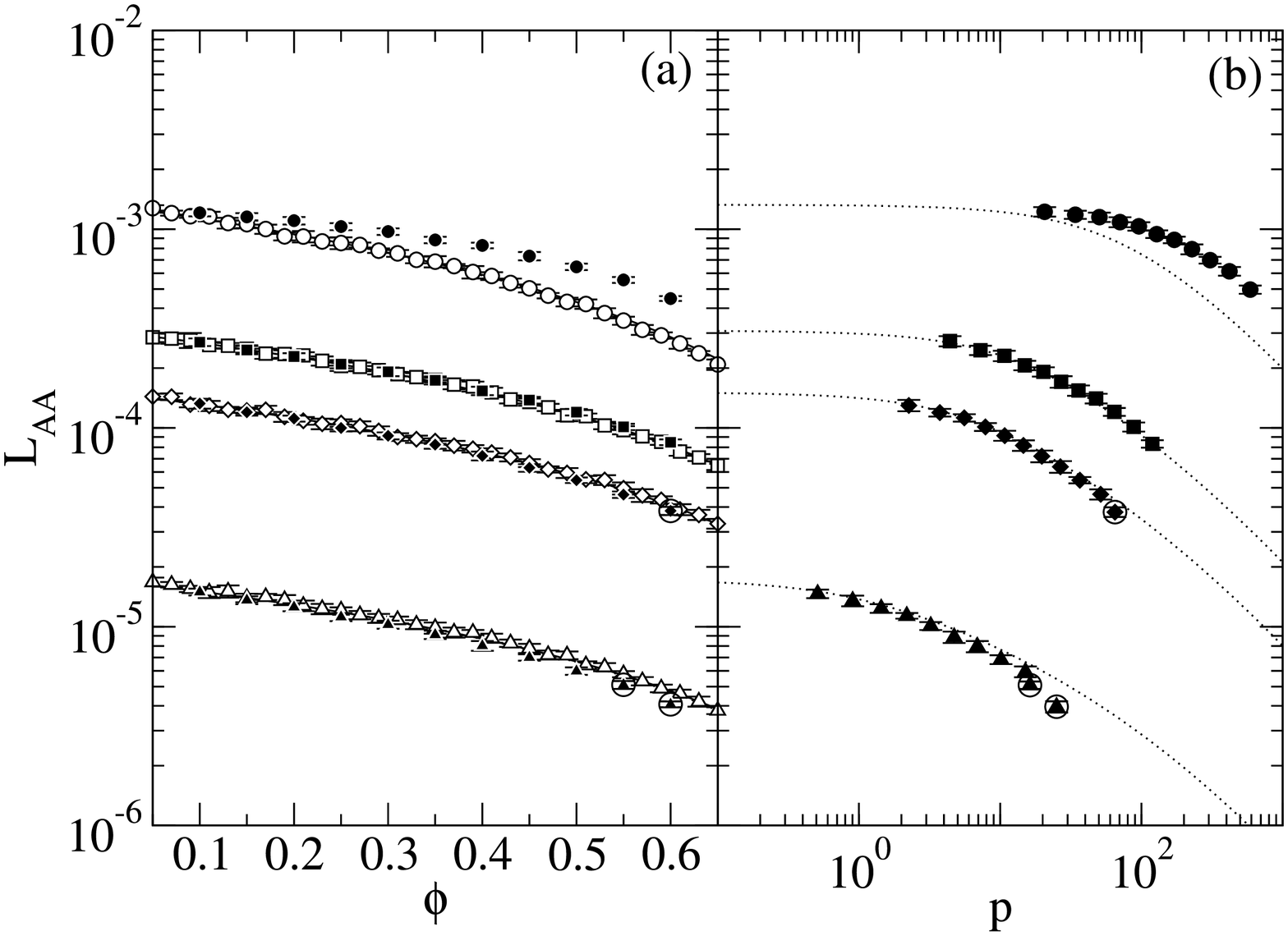}
\caption{\label{fig:mutualdiff} 
Mutual diffusion coefficient $L_{AA}$ as a function of solid fraction
$\phi$ (a) and pressure $p$ (b) for binary hard sphere mixtures with $\sigma_B/\sigma_A=0.1$,
$m_B/m_A=0.001$, and
(i) $x_A=0.01$ (circles),
(ii) $x_A=0.05$ (squares), 
(iii) $x_A=0.1$ (diamonds), and
(iv) $x_A=0.5$ (triangles).
The filled symbols are from molecular dynamics simulations, and the
open symbols are the DSMC results for the Enskog theory.  The lines
are third order Enskog theory predictions using the BMCSL (solid) and
HC2 (dotted) predictions for $g_{ab}(\sigma_{ab}^+)$. Simulation data
points are circled where the system shows signs of freezing.}
\end{center} 
\end{figure}

The mutual diffusion coefficient of the binary hard sphere mixtures is
plotted in Fig.~\ref{fig:mutualdiff}.  The mutual diffusion
coefficient behaves similarly to the thermal diffusivity.
%
%
The displacement functions required to calculated this transport
coefficient contain no potential terms, and therefore, they do not
contain a collisional component of the flux (see
Tables~\ref{tab:displacementfunc} and \ref{tab:coeffs}).
Consequently, Enskog theory performs equally well with HC2 or BMCSL
contact radial distribution values.  Similar to the results for the
thermal diffusivity, Enskog theory is in quantitative agreement with
the simulation data over most of the conditions examined, with the
exception of the $x_A=0.01$ systems, where it significantly
underpredicts the diffusion coefficient.


\section{Conclusions 
\label{sec:conclusions}} 

In this work, we examined the properties of binary mixtures of hard
spheres with a diameter ratio of $\sigma_B/\sigma_A=0.1$ and a mass
ratio of $m_B/m_A=0.001$. The BMCSL equation of state is able to
accurately describe the pressure for all the conditions that we
investigated where the system did not freeze.  However, it
underpredicts the value of $g_{AB}$ and $g_{AA}$, especially at high
packing fractions and low mole fractions of the larger spheres.  The
recently developed HC2 equation, however, is able to quantitatively
predict these quantities.

Enskog theory provides fairly accurate predictions for the transport
coefficients of the systems that we studied in this work.  The third
order Sonine approximation and the DSMC results agree well with one
another, both validating the DSMC code and demonstrating that the
third order solution is sufficiently accurate over the conditions
studied.  At low mole fractions of the larger hard spheres, Enskog
theory fails to capture the behavior of the transport properties,
especially the shear viscosity.  This may be due to the increased
correlations in the collisions between the larger spheres caused by
the depletion forces due to the presence of the smaller spheres.

DSMC provides a speed benefit over traditional molecular dynamics
simulations where large size asymmetries and low mole fractions are
computationally expensive.  Unfortunately, this is where Enskog theory
begins to break down in predicting the transport properties of the
fluid.  Extension of DSMC to other kinetic theories, such as ring
theory, is necessary to capture this behavior, however, these
techniques are yet to be developed.


We find a ``fines'' effect where the addition of smaller spheres to a
larger hard sphere fluid decreases the viscosity of the system, which
occurs at packing fractions greater than about 0.4. This effect is not
captured by Enskog theory. With the addition of fines, the thermal
conductivity of the mixture no longer monotonically increases with the
packing fraction but instead initially decreases with increasing
packing fraction to a minimum value and then increases.  In addition,
at low to moderate packing fractions, there is a region in $x_A$ where
the thermal conductivity of the mixture is higher than thermal
conductivity of either pure species.

\begin{acknowledgments}
MN Bannerman acknowledges support from an EPSRC DTA.
\end{acknowledgments}

\bibliography{main2}

\begin{thebibliography}{52}
\expandafter\ifx\csname natexlab\endcsname\relax\def\natexlab#1{#1}\fi
\expandafter\ifx\csname bibnamefont\endcsname\relax
  \def\bibnamefont#1{#1}\fi
\expandafter\ifx\csname bibfnamefont\endcsname\relax
  \def\bibfnamefont#1{#1}\fi
\expandafter\ifx\csname citenamefont\endcsname\relax
  \def\citenamefont#1{#1}\fi
\expandafter\ifx\csname url\endcsname\relax
  \def\url#1{\texttt{#1}}\fi
\expandafter\ifx\csname urlprefix\endcsname\relax\def\urlprefix{URL }\fi
\providecommand{\bibinfo}[2]{#2}
\providecommand{\eprint}[2][]{\url{#2}}

\bibitem[{\citenamefont{Hansen and McDonald}(1986)}]{HANSEN_MCDONALD_1986}
\bibinfo{author}{\bibfnamefont{J.~P.} \bibnamefont{Hansen}} \bibnamefont{and}
  \bibinfo{author}{\bibfnamefont{I.~R.} \bibnamefont{McDonald}},
  \emph{\bibinfo{title}{Theory of Simple Liquids}}
  (\bibinfo{publisher}{Academic Press}, \bibinfo{address}{London},
  \bibinfo{year}{1986}), \bibinfo{edition}{2nd} ed.

\bibitem[{\citenamefont{Oosawa and Asakura}(1954)}]{Oosawa_Asakura_1954}
\bibinfo{author}{\bibfnamefont{F.}~\bibnamefont{Oosawa}} \bibnamefont{and}
  \bibinfo{author}{\bibfnamefont{S.}~\bibnamefont{Asakura}},
  \bibinfo{journal}{J. Chem. Phys.} \textbf{\bibinfo{volume}{22}},
  \bibinfo{pages}{1255} (\bibinfo{year}{1954}).

\bibitem[{\citenamefont{Vrij}(1976)}]{Vrij_1976}
\bibinfo{author}{\bibfnamefont{A.}~\bibnamefont{Vrij}}, \bibinfo{journal}{Pure
  Appl. Chem.} \textbf{\bibinfo{volume}{48}}, \bibinfo{pages}{471}
  (\bibinfo{year}{1976}).

\bibitem[{\citenamefont{Jackson et~al.}(1987)\citenamefont{Jackson, Rowlinson,
  and van Swol}}]{JACKSON_ETAL_1987}
\bibinfo{author}{\bibfnamefont{G.}~\bibnamefont{Jackson}},
  \bibinfo{author}{\bibfnamefont{J.~S.} \bibnamefont{Rowlinson}},
  \bibnamefont{and} \bibinfo{author}{\bibfnamefont{F.}~\bibnamefont{van Swol}},
  \bibinfo{journal}{J. Phys. Chem} \textbf{\bibinfo{volume}{91}},
  \bibinfo{pages}{4907} (\bibinfo{year}{1987}).

\bibitem[{\citenamefont{Yau et~al.}(1996)\citenamefont{Yau, Chan, and
  Henderson}}]{Yau_etal_1996}
\bibinfo{author}{\bibfnamefont{D.~H.~L.} \bibnamefont{Yau}},
  \bibinfo{author}{\bibfnamefont{K.-Y.} \bibnamefont{Chan}}, \bibnamefont{and}
  \bibinfo{author}{\bibfnamefont{D.}~\bibnamefont{Henderson}},
  \bibinfo{journal}{Mol. Phys.} \textbf{\bibinfo{volume}{88}},
  \bibinfo{pages}{1237} (\bibinfo{year}{1996}).

\bibitem[{\citenamefont{Yau et~al.}(1997)\citenamefont{Yau, Chan, and
  Henderson}}]{Yau_etal_1997}
\bibinfo{author}{\bibfnamefont{D.~H.~L.} \bibnamefont{Yau}},
  \bibinfo{author}{\bibfnamefont{K.-Y.} \bibnamefont{Chan}}, \bibnamefont{and}
  \bibinfo{author}{\bibfnamefont{D.}~\bibnamefont{Henderson}},
  \bibinfo{journal}{Mol. Phys.} \textbf{\bibinfo{volume}{91}},
  \bibinfo{pages}{1137} (\bibinfo{year}{1997}).

\bibitem[{\citenamefont{Henderson et~al.}(1996)\citenamefont{Henderson, Yau,
  and Kwong-Yu}}]{HENDERSON_ETAL_1996_2}
\bibinfo{author}{\bibfnamefont{D.}~\bibnamefont{Henderson}},
  \bibinfo{author}{\bibfnamefont{D.~H.~L.} \bibnamefont{Yau}},
  \bibnamefont{and} \bibinfo{author}{\bibfnamefont{C.}~\bibnamefont{Kwong-Yu}},
  \bibinfo{journal}{Mol. Phys.} \textbf{\bibinfo{volume}{88}},
  \bibinfo{pages}{1237} (\bibinfo{year}{1996}).

\bibitem[{\citenamefont{Lue and Woodcock}(1999)}]{LUE_WOODCOCK_1999}
\bibinfo{author}{\bibfnamefont{L.}~\bibnamefont{Lue}} \bibnamefont{and}
  \bibinfo{author}{\bibfnamefont{L.~V.} \bibnamefont{Woodcock}},
  \bibinfo{journal}{Mol. Phys.} \textbf{\bibinfo{volume}{96}},
  \bibinfo{pages}{1435} (\bibinfo{year}{1999}).

\bibitem[{\citenamefont{Chan and Henderson}(2000)}]{Chan_Henderson_2000}
\bibinfo{author}{\bibfnamefont{K.-Y.} \bibnamefont{Chan}} \bibnamefont{and}
  \bibinfo{author}{\bibfnamefont{D.}~\bibnamefont{Henderson}},
  \bibinfo{journal}{Mol. Phys.} \textbf{\bibinfo{volume}{98}},
  \bibinfo{pages}{1005} (\bibinfo{year}{2000}).

\bibitem[{\citenamefont{Lue and Woodcock}(2002)}]{LUE_WOODCOCK_2002}
\bibinfo{author}{\bibfnamefont{L.}~\bibnamefont{Lue}} \bibnamefont{and}
  \bibinfo{author}{\bibfnamefont{L.~V.} \bibnamefont{Woodcock}},
  \bibinfo{journal}{Int. J. Thermophys.} \textbf{\bibinfo{volume}{23}},
  \bibinfo{pages}{937} (\bibinfo{year}{2002}).

\bibitem[{\citenamefont{Henderson et~al.}(2005)\citenamefont{Henderson,
  Trokhymchuk, Woodcock, and Kwong-Yu}}]{Henderson_etal_2005}
\bibinfo{author}{\bibfnamefont{D.}~\bibnamefont{Henderson}},
  \bibinfo{author}{\bibfnamefont{A.}~\bibnamefont{Trokhymchuk}},
  \bibinfo{author}{\bibfnamefont{L.~V.} \bibnamefont{Woodcock}},
  \bibnamefont{and} \bibinfo{author}{\bibfnamefont{C.}~\bibnamefont{Kwong-Yu}},
  \bibinfo{journal}{Mol. Phys.} \textbf{\bibinfo{volume}{103}},
  \bibinfo{pages}{667} (\bibinfo{year}{2005}).

\bibitem[{\citenamefont{Vrabecz and T\'oth}(2006)}]{Vrabecz_Toacuteth_2006}
\bibinfo{author}{\bibfnamefont{A.}~\bibnamefont{Vrabecz}} \bibnamefont{and}
  \bibinfo{author}{\bibfnamefont{G.}~\bibnamefont{T\'oth}},
  \bibinfo{journal}{Mol. Phys.} \textbf{\bibinfo{volume}{104}},
  \bibinfo{pages}{1843} (\bibinfo{year}{2006}).

\bibitem[{\citenamefont{Alawneh and
  Henderson}(2008{\natexlab{a}})}]{Alawneh_Henderson_2008}
\bibinfo{author}{\bibfnamefont{M.}~\bibnamefont{Alawneh}} \bibnamefont{and}
  \bibinfo{author}{\bibfnamefont{D.}~\bibnamefont{Henderson}},
  \bibinfo{journal}{Mol. Phys.} \textbf{\bibinfo{volume}{106}},
  \bibinfo{pages}{607} (\bibinfo{year}{2008}{\natexlab{a}}).

\bibitem[{\citenamefont{Alawneh and
  Henderson}(2008{\natexlab{b}})}]{Alawneh_Henderson_2008err}
\bibinfo{author}{\bibfnamefont{M.}~\bibnamefont{Alawneh}} \bibnamefont{and}
  \bibinfo{author}{\bibfnamefont{D.}~\bibnamefont{Henderson}},
  \bibinfo{journal}{Mol. Phys.} \textbf{\bibinfo{volume}{106}},
  \bibinfo{pages}{2407} (\bibinfo{year}{2008}{\natexlab{b}}).

\bibitem[{\citenamefont{Santos and L\'opez~de Haro}(2005)}]{Santos_2005}
\bibinfo{author}{\bibfnamefont{A.}~\bibnamefont{Santos}} \bibnamefont{and}
  \bibinfo{author}{\bibfnamefont{M.~L.} \bibnamefont{L\'opez~de Haro}},
  \bibinfo{journal}{J. Chem. Phys.} \textbf{\bibinfo{volume}{123}},
  \bibinfo{pages}{234512} (\bibinfo{year}{2005}).

\bibitem[{\citenamefont{Herman and Alder}(1972)}]{HERMAN_ALDER_1972}
\bibinfo{author}{\bibfnamefont{P.~T.} \bibnamefont{Herman}} \bibnamefont{and}
  \bibinfo{author}{\bibfnamefont{B.~J.} \bibnamefont{Alder}},
  \bibinfo{journal}{J. Chem. Phys.} \textbf{\bibinfo{volume}{56}},
  \bibinfo{pages}{987} (\bibinfo{year}{1972}).

\bibitem[{\citenamefont{Subramanian et~al.}(1974)\citenamefont{Subramanian,
  Levitt, and Davis}}]{LEVITT_DAVIS_1974}
\bibinfo{author}{\bibfnamefont{G.}~\bibnamefont{Subramanian}},
  \bibinfo{author}{\bibfnamefont{D.}~\bibnamefont{Levitt}}, \bibnamefont{and}
  \bibinfo{author}{\bibfnamefont{H.~T.} \bibnamefont{Davis}},
  \bibinfo{journal}{J. Chem. Phys.} \textbf{\bibinfo{volume}{60}},
  \bibinfo{pages}{591} (\bibinfo{year}{1974}).

\bibitem[{\citenamefont{Alder et~al.}(1974)\citenamefont{Alder, Alley, and
  Dymond}}]{ALDER_ALLEY_DYMOND_1974}
\bibinfo{author}{\bibfnamefont{B.~J.} \bibnamefont{Alder}},
  \bibinfo{author}{\bibfnamefont{W.~E.} \bibnamefont{Alley}}, \bibnamefont{and}
  \bibinfo{author}{\bibfnamefont{J.~H.} \bibnamefont{Dymond}},
  \bibinfo{journal}{J. Chem. Phys.} \textbf{\bibinfo{volume}{61}},
  \bibinfo{pages}{1415} (\bibinfo{year}{1974}).

\bibitem[{\citenamefont{Rudyak et~al.}(2000)\citenamefont{Rudyak, Kharlamov,
  and Belkin}}]{RUDYAK_KHARLAMOV_BELKIN_2000}
\bibinfo{author}{\bibfnamefont{V.~Y.} \bibnamefont{Rudyak}},
  \bibinfo{author}{\bibfnamefont{G.~V.} \bibnamefont{Kharlamov}},
  \bibnamefont{and} \bibinfo{author}{\bibfnamefont{A.~A.}
  \bibnamefont{Belkin}}, \bibinfo{journal}{Tech. Phys. Let.}
  \textbf{\bibinfo{volume}{26}}, \bibinfo{pages}{553} (\bibinfo{year}{2000}).

\bibitem[{\citenamefont{Rudyak et~al.}(2001)\citenamefont{Rudyak, Kharlamov,
  and Belkin}}]{RUDYAK_KHARLAMOV_BELKIN_2001}
\bibinfo{author}{\bibfnamefont{V.~Y.} \bibnamefont{Rudyak}},
  \bibinfo{author}{\bibfnamefont{G.~V.} \bibnamefont{Kharlamov}},
  \bibnamefont{and} \bibinfo{author}{\bibfnamefont{A.~A.}
  \bibnamefont{Belkin}}, \bibinfo{journal}{High Temp.}
  \textbf{\bibinfo{volume}{39}}, \bibinfo{pages}{264} (\bibinfo{year}{2001}).

\bibitem[{\citenamefont{Easteal and Woolf}(1990)}]{EASTEAL_WOOLF_1990}
\bibinfo{author}{\bibfnamefont{A.~J.} \bibnamefont{Easteal}} \bibnamefont{and}
  \bibinfo{author}{\bibfnamefont{L.~A.} \bibnamefont{Woolf}},
  \bibinfo{journal}{Chem. Phys. Lett.} \textbf{\bibinfo{volume}{167}},
  \bibinfo{pages}{329} (\bibinfo{year}{1990}).

\bibitem[{\citenamefont{Erpenbeck}(1989)}]{ERPENBECK_1989}
\bibinfo{author}{\bibfnamefont{J.~J.} \bibnamefont{Erpenbeck}},
  \bibinfo{journal}{Phys. Rev. A} \textbf{\bibinfo{volume}{39}},
  \bibinfo{pages}{4718} (\bibinfo{year}{1989}).

\bibitem[{\citenamefont{Erpenbeck}(1992)}]{ERPENBECK_1992}
\bibinfo{author}{\bibfnamefont{J.~J.} \bibnamefont{Erpenbeck}},
  \bibinfo{journal}{Phys. Rev. A} \textbf{\bibinfo{volume}{45}},
  \bibinfo{pages}{2298} (\bibinfo{year}{1992}).

\bibitem[{\citenamefont{Erpenbeck}(1993)}]{ERPENBECK_1993}
\bibinfo{author}{\bibfnamefont{J.~J.} \bibnamefont{Erpenbeck}},
  \bibinfo{journal}{Phys. Rev. E} \textbf{\bibinfo{volume}{48}},
  \bibinfo{pages}{223} (\bibinfo{year}{1993}).

\bibitem[{\citenamefont{Yeganegi and
  Zolfaghari}(2006)}]{Yeganegi_Zolfaghari_2006}
\bibinfo{author}{\bibfnamefont{S.}~\bibnamefont{Yeganegi}} \bibnamefont{and}
  \bibinfo{author}{\bibfnamefont{M.}~\bibnamefont{Zolfaghari}},
  \bibinfo{journal}{Fluid Phase Equilibria} \textbf{\bibinfo{volume}{243}},
  \bibinfo{pages}{161} (\bibinfo{year}{2006}), ISSN
  \bibinfo{issn}{{0378-3812}}.

\bibitem[{\citenamefont{Bastea}(2007)}]{Bastea_2007}
\bibinfo{author}{\bibfnamefont{S.}~\bibnamefont{Bastea}},
  \bibinfo{journal}{Phys. Rev. E} \textbf{\bibinfo{volume}{75}},
  \bibinfo{pages}{031201} (\bibinfo{year}{2007}).

\bibitem[{\citenamefont{L\'opez~de Haro et~al.}(1983)\citenamefont{L\'opez~de
  Haro, Cohen, and Kincaid}}]{DEHARO_ETAL_1983}
\bibinfo{author}{\bibfnamefont{M.}~\bibnamefont{L\'opez~de Haro}},
  \bibinfo{author}{\bibfnamefont{E.~G.~D.} \bibnamefont{Cohen}},
  \bibnamefont{and} \bibinfo{author}{\bibfnamefont{J.~M.}
  \bibnamefont{Kincaid}}, \bibinfo{journal}{J. Chem. Phys.}
  \textbf{\bibinfo{volume}{78}}, \bibinfo{pages}{2746} (\bibinfo{year}{1983}).

\bibitem[{\citenamefont{Matyushov and Ladanyi}(1997)}]{Matyushov_Ladanyi_1997}
\bibinfo{author}{\bibfnamefont{D.~V.} \bibnamefont{Matyushov}}
  \bibnamefont{and} \bibinfo{author}{\bibfnamefont{B.}~\bibnamefont{Ladanyi}},
  \bibinfo{journal}{J. Chem. Phys.} \textbf{\bibinfo{volume}{107}},
  \bibinfo{pages}{5815} (\bibinfo{year}{1997}).

\bibitem[{\citenamefont{Boublik}(1970)}]{Boublik_1970}
\bibinfo{author}{\bibfnamefont{T.}~\bibnamefont{Boublik}}, \bibinfo{journal}{J.
  Chem. Phys.} \textbf{\bibinfo{volume}{53}}, \bibinfo{pages}{471}
  (\bibinfo{year}{1970}).

\bibitem[{\citenamefont{Mansoori et~al.}(1971)\citenamefont{Mansoori, Carnahan,
  Starling, and Leland}}]{MANSOORI_CARNAHAN_STARLING_1971}
\bibinfo{author}{\bibfnamefont{G.~A.} \bibnamefont{Mansoori}},
  \bibinfo{author}{\bibfnamefont{N.~F.} \bibnamefont{Carnahan}},
  \bibinfo{author}{\bibfnamefont{K.~E.} \bibnamefont{Starling}},
  \bibnamefont{and} \bibinfo{author}{\bibfnamefont{T.~W.}
  \bibnamefont{Leland}}, \bibinfo{journal}{J. Chem. Phys.}
  \textbf{\bibinfo{volume}{54}}, \bibinfo{pages}{1523} (\bibinfo{year}{1971}).

\bibitem[{\citenamefont{Percus and Yevick}(1958)}]{PERCUS_YEVICK_1958}
\bibinfo{author}{\bibfnamefont{J.~K.} \bibnamefont{Percus}} \bibnamefont{and}
  \bibinfo{author}{\bibfnamefont{G.~J.} \bibnamefont{Yevick}},
  \bibinfo{journal}{Phys. Rev.} \textbf{\bibinfo{volume}{110}},
  \bibinfo{pages}{1} (\bibinfo{year}{1958}).

\bibitem[{\citenamefont{Roth et~al.}(2000)\citenamefont{Roth, Evans, and
  Dietrich}}]{Roth_etal_2000}
\bibinfo{author}{\bibfnamefont{R.}~\bibnamefont{Roth}},
  \bibinfo{author}{\bibfnamefont{R.}~\bibnamefont{Evans}}, \bibnamefont{and}
  \bibinfo{author}{\bibfnamefont{S.}~\bibnamefont{Dietrich}},
  \bibinfo{journal}{Phys. Rev. E} \textbf{\bibinfo{volume}{57}},
  \bibinfo{pages}{6785} (\bibinfo{year}{2000}).

\bibitem[{\citenamefont{Viduna and
  Smith}(2002{\natexlab{a}})}]{Viduna_Smith_2002}
\bibinfo{author}{\bibfnamefont{D.}~\bibnamefont{Viduna}} \bibnamefont{and}
  \bibinfo{author}{\bibfnamefont{W.~R.} \bibnamefont{Smith}},
  \bibinfo{journal}{J. Chem. Phys.} \textbf{\bibinfo{volume}{117}},
  \bibinfo{pages}{1214} (\bibinfo{year}{2002}{\natexlab{a}}).

\bibitem[{\citenamefont{Viduna and
  Smith}(2002{\natexlab{b}})}]{Viduna_Smith_2002_2}
\bibinfo{author}{\bibfnamefont{D.}~\bibnamefont{Viduna}} \bibnamefont{and}
  \bibinfo{author}{\bibfnamefont{W.~R.} \bibnamefont{Smith}},
  \bibinfo{journal}{Mol. Phys.} \textbf{\bibinfo{volume}{100}},
  \bibinfo{pages}{2903} (\bibinfo{year}{2002}{\natexlab{b}}).

\bibitem[{\citenamefont{de~Groot and Mazur}(1984)}]{GROOT_MAZUR_1984}
\bibinfo{author}{\bibfnamefont{S.}~\bibnamefont{de~Groot}} \bibnamefont{and}
  \bibinfo{author}{\bibfnamefont{P.}~\bibnamefont{Mazur}},
  \emph{\bibinfo{title}{Non-Equilibrium Thermodynamics}}
  (\bibinfo{publisher}{Dover}, \bibinfo{address}{New York},
  \bibinfo{year}{1984}).

\bibitem[{\citenamefont{Kincaid et~al.}(1983)\citenamefont{Kincaid, L\'opez~de
  Haro, and Cohen}}]{KINCAID_ETAL_1983}
\bibinfo{author}{\bibfnamefont{J.~M.} \bibnamefont{Kincaid}},
  \bibinfo{author}{\bibfnamefont{M.}~\bibnamefont{L\'opez~de Haro}},
  \bibnamefont{and} \bibinfo{author}{\bibfnamefont{E.~G.~D.}
  \bibnamefont{Cohen}}, \bibinfo{journal}{J. Chem. Phys.}
  \textbf{\bibinfo{volume}{79}}, \bibinfo{pages}{4509} (\bibinfo{year}{1983}).

\bibitem[{\citenamefont{L\'opez~de Haro and Cohen}(1984)}]{DEHARO_COHEN_1984}
\bibinfo{author}{\bibfnamefont{M.~L.} \bibnamefont{L\'opez~de Haro}}
  \bibnamefont{and} \bibinfo{author}{\bibfnamefont{E.~G.~D.}
  \bibnamefont{Cohen}}, \bibinfo{journal}{J. Chem. Phys.}
  \textbf{\bibinfo{volume}{80}}, \bibinfo{pages}{408} (\bibinfo{year}{1984}).

\bibitem[{\citenamefont{Kincaid et~al.}(1987)\citenamefont{Kincaid, Cohen, and
  L\'opez~de Haro}}]{KINCAID_ETAL_1987}
\bibinfo{author}{\bibfnamefont{J.~M.} \bibnamefont{Kincaid}},
  \bibinfo{author}{\bibfnamefont{E.~G.~D.} \bibnamefont{Cohen}},
  \bibnamefont{and} \bibinfo{author}{\bibfnamefont{M.}~\bibnamefont{L\'opez~de
  Haro}}, \bibinfo{journal}{J. Chem. Phys.} \textbf{\bibinfo{volume}{86}},
  \bibinfo{pages}{963} (\bibinfo{year}{1987}).

\bibitem[{\citenamefont{Chapman and Cowling}(1970)}]{CHAPMAN_COWLING_1970}
\bibinfo{author}{\bibfnamefont{S.}~\bibnamefont{Chapman}} \bibnamefont{and}
  \bibinfo{author}{\bibfnamefont{T.~G.} \bibnamefont{Cowling}},
  \emph{\bibinfo{title}{The {M}athematical {T}heory of {N}on-{U}niform
  {G}ases}} (\bibinfo{publisher}{Cambridge}, \bibinfo{address}{Cambridge},
  \bibinfo{year}{1970}), \bibinfo{edition}{3rd} ed.

\bibitem[{\citenamefont{Ferziger and Kaper}(1972)}]{FERZIGER_KAPER_1972}
\bibinfo{author}{\bibfnamefont{J.~H.} \bibnamefont{Ferziger}} \bibnamefont{and}
  \bibinfo{author}{\bibfnamefont{H.~G.} \bibnamefont{Kaper}},
  \emph{\bibinfo{title}{Mathematical Theory of Transport Processes in Gases}}
  (\bibinfo{publisher}{North-Holland}, \bibinfo{address}{London},
  \bibinfo{year}{1972}).

\bibitem[{\citenamefont{Montanero and Santos}(1996)}]{Montanero1996a}
\bibinfo{author}{\bibfnamefont{J.~M.} \bibnamefont{Montanero}}
  \bibnamefont{and} \bibinfo{author}{\bibfnamefont{A.}~\bibnamefont{Santos}},
  \bibinfo{journal}{Phys. Rev. E} \textbf{\bibinfo{volume}{54}},
  \bibinfo{pages}{438} (\bibinfo{year}{1996}).

\bibitem[{\citenamefont{Montanero and Santos}(1997)}]{Montanero1997}
\bibinfo{author}{\bibfnamefont{J.~M.} \bibnamefont{Montanero}}
  \bibnamefont{and} \bibinfo{author}{\bibfnamefont{A.}~\bibnamefont{Santos}},
  \bibinfo{journal}{Phys. Fluids} \textbf{\bibinfo{volume}{9}},
  \bibinfo{pages}{2057} (\bibinfo{year}{1997}).

\bibitem[{\citenamefont{Frezzotti}(1997)}]{Frezzotti_1997}
\bibinfo{author}{\bibfnamefont{A.}~\bibnamefont{Frezzotti}},
  \bibinfo{journal}{Phys. Fluids} \textbf{\bibinfo{volume}{9}},
  \bibinfo{pages}{1329} (\bibinfo{year}{1997}).

\bibitem[{\citenamefont{Bird}(1994)}]{bird_1994}
\bibinfo{author}{\bibfnamefont{G.~A.} \bibnamefont{Bird}},
  \emph{\bibinfo{title}{Molecular gas dynamics and the direct simulation of gas
  flows}} (\bibinfo{publisher}{Oxford Science}, \bibinfo{year}{1994}).

\bibitem[{\citenamefont{Hopkins and Shen}(1992)}]{Hopkins_Shen_1992}
\bibinfo{author}{\bibfnamefont{M.~A.} \bibnamefont{Hopkins}} \bibnamefont{and}
  \bibinfo{author}{\bibfnamefont{H.~H.} \bibnamefont{Shen}},
  \bibinfo{journal}{J. Fluid Mech.} \textbf{\bibinfo{volume}{244}},
  \bibinfo{pages}{477} (\bibinfo{year}{1992}).

\bibitem[{\citenamefont{Alder and Wainwright}(1959)}]{ALDER_WAINWRIGHT_1959}
\bibinfo{author}{\bibfnamefont{B.~J.} \bibnamefont{Alder}} \bibnamefont{and}
  \bibinfo{author}{\bibfnamefont{T.~E.} \bibnamefont{Wainwright}},
  \bibinfo{journal}{J. Chem. Phys.} \textbf{\bibinfo{volume}{31}},
  \bibinfo{pages}{459} (\bibinfo{year}{1959}).

\bibitem[{\citenamefont{Marin et~al.}(1993)\citenamefont{Marin, Risso, and
  Cordero}}]{Marin_etal_1993}
\bibinfo{author}{\bibfnamefont{M.}~\bibnamefont{Marin}},
  \bibinfo{author}{\bibfnamefont{D.}~\bibnamefont{Risso}}, \bibnamefont{and}
  \bibinfo{author}{\bibfnamefont{P.}~\bibnamefont{Cordero}},
  \bibinfo{journal}{J. Comput. Phys.} \textbf{\bibinfo{volume}{109}},
  \bibinfo{pages}{306} (\bibinfo{year}{1993}).

\bibitem[{\citenamefont{Paul}(2007)}]{Paul_2007}
\bibinfo{author}{\bibfnamefont{G.}~\bibnamefont{Paul}}, \bibinfo{journal}{J.
  Comp. Phys.} \textbf{\bibinfo{volume}{221}}, \bibinfo{pages}{615}
  (\bibinfo{year}{2007}).

\bibitem[{\citenamefont{Haile}(1997)}]{Haile_1997}
\bibinfo{author}{\bibfnamefont{J.~M.} \bibnamefont{Haile}},
  \emph{\bibinfo{title}{Molecular Dynamics Simulation - Elementary Methods}}
  (\bibinfo{publisher}{Wiley-Interscience}, \bibinfo{address}{New York},
  \bibinfo{year}{1997}).

\bibitem[{\citenamefont{Hoover and Ree}(1968)}]{Hoover_Ree_1968}
\bibinfo{author}{\bibfnamefont{W.~G.} \bibnamefont{Hoover}} \bibnamefont{and}
  \bibinfo{author}{\bibfnamefont{F.~H.} \bibnamefont{Ree}},
  \bibinfo{journal}{J. Chem. Phys.} \textbf{\bibinfo{volume}{49}},
  \bibinfo{pages}{3609} (\bibinfo{year}{1968}).

\bibitem[{\citenamefont{Lue}(2005)}]{Lue_2005}
\bibinfo{author}{\bibfnamefont{L.}~\bibnamefont{Lue}}, \bibinfo{journal}{J.
  Chem. Phys.} \textbf{\bibinfo{volume}{122}}, \bibinfo{pages}{044513}
  (\bibinfo{year}{2005}).

\bibitem[{\citenamefont{Lue and Bishop}(2006)}]{Lue_Bishop_2006}
\bibinfo{author}{\bibfnamefont{L.}~\bibnamefont{Lue}} \bibnamefont{and}
  \bibinfo{author}{\bibfnamefont{M.}~\bibnamefont{Bishop}},
  \bibinfo{journal}{Phys. Rev. E} \textbf{\bibinfo{volume}{74}},
  \bibinfo{pages}{021201} (\bibinfo{year}{2006}).

\end{thebibliography}
\bibliographystyle{apsrev}

\clearpage

\begin{table}
\begin{center}
\caption{\label{tab:displacementfunc} 
  Displacement functions for an isotropic system required to evaluate the
  Einstein form of the Green-Kubo relationships, see
  Eq.~(\ref{eq:Einstein}).  }
\begin{tabular*}{0.75\columnwidth}{@{\extracolsep{\fill}}ll}
  \hline\hline
  $W_\psi$ & \\
  \hline
  ${\bf W}_a$ & $\sum_{\Delta t_c}^{t}\sum_{k}^{N_a}m_k{\bf v}_k\Delta t_c -
  c_a \sum_{k}^{N}m_k{\bf v}_k\Delta t_c$ \\
  ${\bf W}_\lambda$ & $\sum_{\Delta t_c}^{t}
    \left(\sum_{k}^{N}\frac{1}{2} m_k v_k^2 {\bf v}_k \Delta t_c
    + \frac{1}{2}m_i\Delta v_i^2 {\bf v}_{ij} \right)$ \\
  ${\bf W}_\eta$ & $\sum_{\Delta t_c}^{t}\left(\sum_{k}^{N}m_k{\bf v}_k{\bf v}_{k}\Delta
    t_c + m_i {\bf r}_{ij} \Delta {\bf v}_{i}  - {\bf 1} pV \Delta t_c\right)$\\
  \hline\hline
\end{tabular*}
\footnotetext{The first summation runs over all time intervals between
  collisions $\Delta t_c$ that occur during the simulation time $t$.
  The indexes $i$ and $j$ denote the pair of spheres that undergo
  collision at the end of this time interval.  Note that $c_a$ is the
  mass fraction of sphere of type $a$.  }
\end{center}
\end{table}

\clearpage

\begin{table}
\begin{center}  
\caption{\label{tab:coeffs} 
  Transport coefficients and the corresponding displacement functions.
  The right hand columns indicate which rows of
  Table~\ref{tab:displacementfunc} are used.}
  \begin{tabular*}{0.75\columnwidth}{@{\extracolsep{\fill}}lll}
    \hline\hline
    $\psi$ & $W_{\psi_1}$ & $W_{\psi_2}$\\
    \hline
    $L_{ab}$ & $W_{a,x}$ & $W_{b,x}$ \\
    $L_{a\lambda}$ & $W_{a,x}$ & $W_{\lambda,x}$ \\
    $L_{\lambda\lambda}$ & $W_{\lambda,x}$ & $W_{\lambda,x}$ \\
    $\eta$ & $W_{\eta,xy}$ & $W_{\eta,xy}$\\
    $\frac{4}{3}\eta + \kappa$ & $W_{\eta,xx}$ & $W_{\eta,xx}$ \\
    \hline\hline
  \end{tabular*}
  \footnotetext{As the system is isotropic, the transport coefficients
    are averaged over all components $x\neq y$ of the
    displacement functions.}
\end{center}
\end{table}

\clearpage

\clearpage

\clearpage

\clearpage

\clearpage

\clearpage

\clearpage

\end{document}